\documentclass[conference]{IEEEtran}
\usepackage{cite}
\usepackage{amsmath,amssymb,amsfonts}
\usepackage{algorithmic}
\usepackage{graphicx}
\usepackage{textcomp}
\usepackage{xcolor}
\usepackage{url}
\usepackage{booktabs}

\def\BibTeX{{\rm B\kern-.05em{\sc i\kern-.025em b}\kern-.08em
    T\kern-.1667em\lower.7ex\hbox{E}\kern-.125emX}}

\begin{document}

\title{Cross-Layer Decision Timing Orchestration in Cost-Based Database Systems: Resolving Structural Temporal Misalignment}

\author{\IEEEauthorblockN{Il-Sun Chang}
\IEEEauthorblockA{\textit{Independent Researcher} \\
Daegu, Republic of Korea \\
kyou0072@gmail.com}
}

\maketitle

\begin{abstract}
This paper analyzes execution instability in traditional cost-based database management systems (DBMS) and identifies a recurring timing misalignment between optimization and execution stages that contributes to tail latency amplification. We argue that, beyond limitations of estimation accuracy and execution throughput, decision timing and the availability of runtime signals can materially affect tail-latency behavior. In traditional DBMS architectures, the optimizer relies on historical statistics, the executor observes runtime data distribution and resource state, and accelerators impose up-front costs and amortization constraints. This temporal asynchrony results in suboptimal early-binding decisions that may fail to adapt to runtime conditions. To address this, we propose a Cross-Layer Decision Timing Orchestration framework that shifts the final decision authority from the compile-time optimizer to the runtime executor through a Late Binding mechanism. We define the Unified Risk Signal (URS), a vector integrating risk metrics from the optimizer, executor, and accelerator, to guide runtime decisions. Experiments on a modified PostgreSQL prototype evaluate (i) input-scale shift, (ii) stale-statistics drift, and (iii) GPU offload break-even regimes using controlled microbenchmarks. The proposed orchestration improves execution stability, reducing P99 latency by up to 20$\times$ under severe estimation drift and unexpected input-scale shifts without degrading median latency.
\end{abstract}


\section{Introduction}

Cost-Based Optimization (CBO) has been the cornerstone of query processing in database systems for decades. The optimizer evaluates alternative execution plans using statistical summaries and cost models to select the plan with the lowest estimated cost. While effective in relatively stable environments, modern systems increasingly exhibit performance instability, plan regression, and tail-latency amplification despite improvements in hardware.

Prior work has addressed these problems through improved statistics, learning-based estimation, adaptive execution, re-optimization, and hardware acceleration. However, most approaches emphasize improving estimation accuracy or execution efficiency. Comparatively less attention has been given to the structural question of when decisions should be finalized.

We observe that traditional DBMS architectures rely on early plan binding based on historical statistics. In contrast, the executor operates on runtime data distributions and resource states, while accelerators impose up-front transfer costs and amortization constraints tied to future workload characteristics. These layers therefore operate on distinct temporal coordinates. A single early-bound decision applied across all phases can lead to instability when runtime conditions diverge from compile-time assumptions.

This paper frames such instability as a structural temporal misalignment problem and proposes a cross-layer orchestration mechanism that relocates final decision authority to the execution phase.

Our prototype does not introduce a new optimizer nor perform full mid-query re-planning. Instead, it selectively late-binds a small set of high-impact operator-level choices (e.g., CPU vs. GPU primitives, join strategy family) using runtime-observable signals. This design explicitly limits overhead and avoids repeated optimization cycles, while targeting stability of tail latency under uncertainty.

\section{Related Work}

\subsection{Adaptive Plans in Commercial Systems}
Commercial systems such as Oracle and Microsoft SQL Server incorporate adaptive query mechanisms that defer certain operator decisions until runtime. For example, adaptive joins in SQL Server \cite{b13} postpone the selection between Nested Loop and Hash Join until the cardinality of the build input is observed. These mechanisms demonstrate the value of limited runtime flexibility but are typically implemented as localized heuristics rather than as an explicit cross-layer orchestration principle.
In contrast to adaptive joins that defer a single local decision (e.g., NLJ vs. hash join) based on an observed build cardinality, we treat decision timing as a cross-layer coordination variable spanning optimizer uncertainty, executor resource state, and accelerator amortization constraints. Our contribution is not another heuristic, but a structured orchestration interface that exposes when and where a decision is finalized.

\subsection{Mid-query and Progressive Re-optimization}
Mid-query re-optimization \cite{b8} detects cardinality estimation errors during execution and triggers re-planning. Progressive optimization \cite{b11} refines this approach by minimizing redundant work during re-optimization. Our framework builds upon this lineage but emphasizes continuous runtime authority rather than interruption-based re-planning.

\subsection{Continuous Adaptivity}
Eddies \cite{b7} propose tuple-level routing for continuous adaptivity. More recent systems such as SkinnerDB \cite{b14} explore reinforcement learning for adaptive join ordering. While highly adaptive, these systems may incur non-trivial overhead in high-throughput environments. Our approach aims for a coarser-grained but system-oriented runtime authority shift guided by integrated risk signals.

\subsection{Parametric Optimization}
Parametric Query Optimization \cite{b9} generates multiple plans at compile time and selects among them at runtime. Although this mitigates some early-binding limitations, it faces scalability challenges as parameter dimensionality increases. Our approach instead integrates runtime signals with accelerator cost constraints dynamically.

\subsection{Distributed Runtime Adaptation}
Distributed engines such as Apache Spark employ Adaptive Query Execution (AQE) \cite{b15}, dynamically adjusting join strategies and shuffle partitioning at runtime. These systems reinforce the practical importance of runtime plan refinement. Our work differs in focusing on kernel-level architectural timing misalignment and incorporating accelerator amortization constraints as explicit risk signals.

\subsection{Differentiation}
Compared to existing approaches, our framework contributes:
\begin{itemize}
    \item \textbf{Structural Framing:} Explicit identification of decision timing misalignment as an architectural factor contributing to instability.
    \item \textbf{Decision Authority Shift:} A design principle relocating final decision authority to the execution layer.
    \item \textbf{Accelerator-aware Risk Integration:} Equal treatment of accelerator up-front costs and amortization requirements as runtime decision signals.
\end{itemize}

\section{The Core Problem: Structural Temporal Misalignment}

\subsection{Temporal Coordinates of DBMS Layers}
Traditional cost-based DBMS consists of different layers: the optimizer, executor, and accelerator, and each layer processes information from fundamentally different time perspectives.
\begin{enumerate}
    \item \textbf{The Optimizer (historical statistics):} The optimizer generates execution plans using statistical information and cost models. The statistics used in this process are summaries of data distribution at a past point in time, and there is an inevitable time gap between the plan generation time and the actual execution time. Therefore, the optimizer's judgment relies on historical statistics.
    \item \textbf{The Executor (runtime data distribution and resource state):} The executor processes actual data tuples during query execution and can directly observe runtime data distribution and resource state at the current time. Phenomena such as a sudden increase in input size observed during execution, explosion of intermediate results according to join order, and memory pressure are information revealed only at the execution stage, which is difficult for the optimizer to predict accurately in advance.
    \item \textbf{The Accelerator (up-front costs and amortization constraints):} Accelerators have another time coordinate. Accelerators like GPUs require upfront costs such as data transfer, initialization, and kernel preparation before execution, and these costs occur before the execution results appear. In other words, accelerators impose up-front costs and amortization constraints that require consideration of future execution patterns. If sufficient workload is not guaranteed, the use of accelerators can rather lead to performance degradation.
\end{enumerate}

In summary, each layer optimizes with signals observed at different stages (compile-time statistics, runtime feedback, and accelerator overhead/amortization), yet a single early-bound plan is typically enforced end-to-end. Existing cost-based systems therefore subordinate the entire execution process to a decision made at a single point in time without explicitly orchestrating these differences in time coordinates.

\subsection{Structural Limitations of Single-Point Decision}
The cost-based optimization structure where decisions are fixed at a single point causes various failure modes. In intervals where statistical information is unreliable, the optimizer selects the wrong execution plan, and if the data scale during execution differs significantly from expectations, the selected plan becomes rapidly inefficient. Also, in the case of accelerators, offloading decisions without securing sufficient workload in the early stages of execution only increase transfer costs and fail to provide substantial performance benefits.

These phenomena have been treated as different problems in existing research. For instance, the problem of inaccurate cardinality estimation has been addressed through more sophisticated statistics or learning-based models, while execution-stage inefficiency has been mitigated through re-optimization or adaptive execution techniques. Accelerator utilization has also been approached as a performance improvement problem limited to specific operations or workloads.

We argue that these issues share a common structural cause: decisions are made too early and enforced too rigidly. In other words, the core problem lies in the timing and location of the decision, prior to estimation accuracy or raw execution throughput.

\section{Proposed Architecture: Decision Timing Orchestration}

\subsection{Decision Authority Shift: From Optimizer-Centric to Executor-Centric}
This study finds the cause of this problem in the question of "who decides when". In existing DBMS, the optimizer monopolizes decision authority, and the executor is treated as a passive component that carries it out. However, the executor is the only layer that can directly observe the actual data flow and resource status. Accordingly, this study redefines the decision structure as follows:
\begin{itemize}
    \item \textbf{Optimizer:} Proposes possible execution strategies, rather than fixing the final execution plan.
    \item \textbf{Executor:} Re-evaluates the proposed decisions based on information observed during execution and performs the final selection.
    \item \textbf{Accelerator:} Provides constraint conditions and future cost information for execution choices.
\end{itemize}
We refer to this shift in execution-time evaluation responsibility as a \textbf{Decision Authority Shift}. In our prototype, \textbf{late binding} applies to operator-level choices such as (i) switching between CPU and GPU execution for selected primitives, and (ii) re-selecting join strategies when runtime input cardinalities deviate beyond a threshold.

\begin{figure}[htbp]
\centerline{\includegraphics[width=\linewidth]{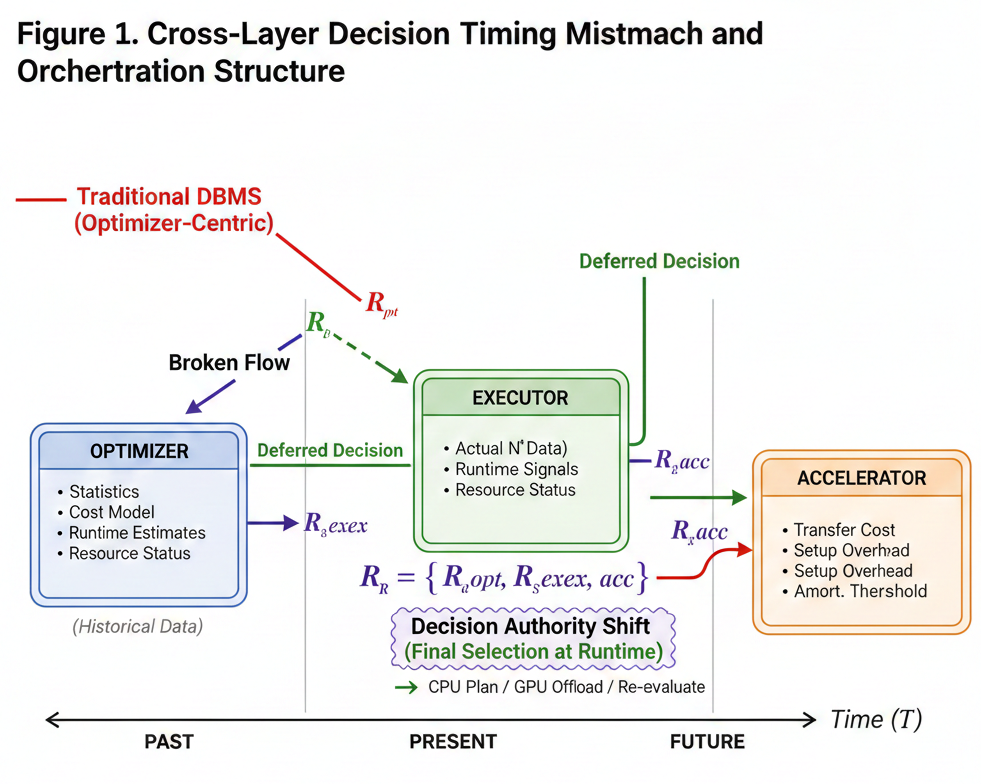}}
\caption{Cross-Layer Decision Timing Mismatch and Orchestration Structure. This figure visualizes the temporal misalignment and the proposed shift in decision authority.}
\label{fig:arch}
\end{figure}

\subsection{Unified Risk Signal (URS)}
To make rational decisions at execution time, an integrated interpretation of risk signals between layers, rather than a single indicator, is required. For this purpose, this study defines the Unified Risk Vector $\mathbf{R}$:
\begin{equation}
\mathbf{R} = (R_{opt}, R_{exec}, R_{acc}).
\end{equation}
Each component has the following meaning:
\begin{itemize}
    \item $R_{opt}$: Risk signal from the optimizer's perspective (e.g., cardinality estimation variance, NDV drift, correlation violation indicators).
    \item $R_{exec}$: Risk signal from the executor's perspective (e.g., surge in actual input size $N$, memory pressure, execution time deviation).
    \item $R_{acc}$: Risk signal from the accelerator's perspective (e.g., data transfer cost, setup overhead, cost amortization threshold).
\end{itemize}
The important point is that these signals are observed at different points in time. Optimizer signals exist only before execution, executor signals are observable only during execution, and accelerator signals include costs after execution. Therefore, integrating them into a single static cost model is difficult in practice due to their differing observation times. Our approach uses these signals as a basis for conditional judgment at runtime, not to reduce them to a single scalar cost.
Formally, URS preserves partial observability across time by keeping components separable until the execution boundary, where conditional policies can be applied without requiring a globally consistent cost model.

\subsection{Execution Time-Based Orchestration Flow}
The execution flow in the proposed structure is as follows:
\begin{enumerate}
    \item The optimizer produces a plan as usual, but allows selected operator-level choices to be finalized at execution time.
    \item The executor observes the $R_{exec}$ signal during actual execution and evaluates $R_{opt}$ and $R_{acc}$ information together if necessary.
    \item If the risk signal satisfies a specific threshold condition, the executor selects one of CPU execution maintenance, GPU offloading, or re-evaluation.
    \item This selection is made dynamically during execution, and the change of prior decision itself is not considered a system error.
\end{enumerate}
In our prototype, decision thresholds were calibrated empirically based on controlled microbenchmarks to approximate break-even points between alternative operator strategies. The key in this structure is not performance optimization, but the appropriateness of the decision timing. What is more important than fast execution is the structural margin to correct decisions made at the wrong time in the execution stage.
Unlike mid-query re-optimization approaches that trigger full plan re-compilation, our orchestration maintains structural continuity by switching only pre-enumerated strategy variants, thereby bounding overhead and preserving execution locality.

\section{Experimental Evaluation and Analysis}
This chapter experimentally verifies how the proposed execution-time-centric cross-layer decision orchestration structure mitigates the structural limitations of existing cost-based database systems. The purpose of the experiment is not to measure the performance improvement of individual operations, but to evaluate the impact of the shift in decision timing and cross-layer orchestration on the stability and decision quality of the entire system.

\subsection{Experimental Setup}
The experiments compare the following system configurations:
\begin{itemize}
    \item \textbf{Baseline (Optimizer-Centric):} Traditional cost-based DBMS structure where the execution plan is fixed at the optimization stage and does not change during execution.
    \item \textbf{Independent Gates:} A variant that applies separate, local gating rules at the optimizer and executor stages without a unified cross-layer signal, approximating common heuristic-based adaptive mechanisms.
    \item \textbf{Proposed Cross-Layer Orchestration:} The structure proposed in this paper, where the executor makes the final decision based on the Unified Risk Signal.
\end{itemize}

\noindent\textbf{Implementation note.} We implement late binding by (i) instrumenting the executor to expose runtime signals (observed input cardinality, memory pressure proxies), and (ii) adding a lightweight decision hook that can switch among pre-enumerated strategy variants at operator boundaries. The optimizer still produces a standard plan, but selected nodes are annotated as late-bind candidates with alternative implementations.

\subsection{Exp 1: Robustness to Input-Scale Shift}
This experiment assumes a situation where the input data scale during execution differs significantly from the optimizer's estimate.
In the Baseline system, since the execution plan is fixed in advance, the selected execution path is maintained even if the input scale increases rapidly during execution. This results in excessive intermediate result generation, memory pressure, and a sharp increase in execution time for some queries. This phenomenon is prominent in the tail of the distribution (P95, P99) rather than the average execution time.

\begin{figure}[htbp]
\centerline{\includegraphics[width=\linewidth]{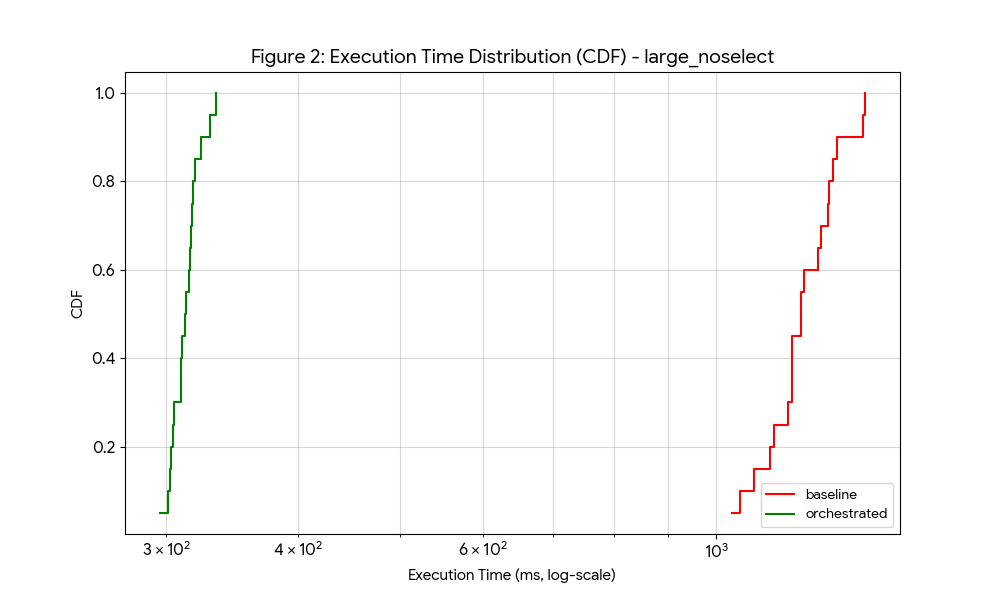}}
\caption{Exp 1: Execution Time Distribution (CDF) - large\_noselect. The Orchestrated system maintains a tight distribution compared to the Baseline.}
\label{fig:exp1_cdf}
\end{figure}

In contrast, in the proposed structure, the executor re-evaluates the pre-determined execution strategy after observing the increase in actual input size. As a result, while maintaining a level similar to the Baseline in average execution time, it showed a tendency to consistently reduce tail latency in the input scale surge interval.

\begin{figure}[htbp]
\centerline{\includegraphics[width=\linewidth]{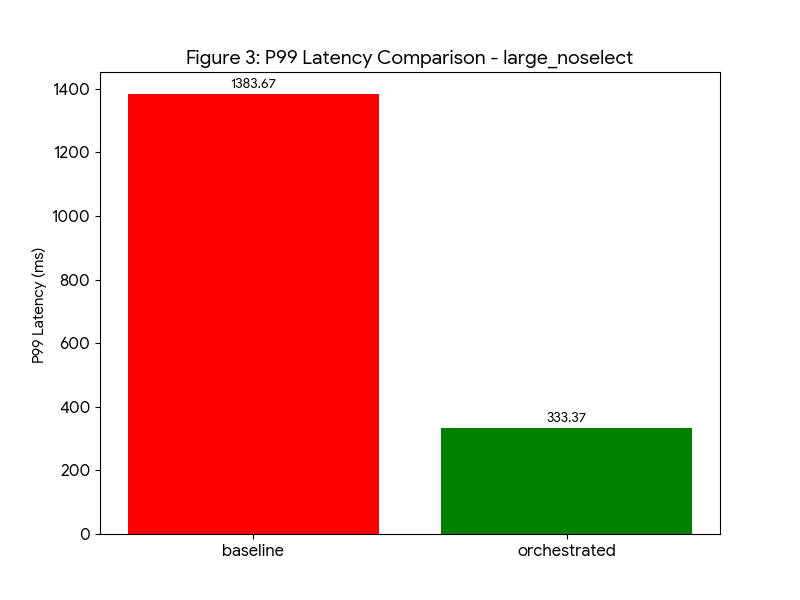}}
\caption{Exp 1: P99 Latency Comparison - large\_noselect.}
\label{fig:exp1_p99}
\end{figure}

\subsection{Exp 2: Resilience Against Stale Statistics}
The second experiment evaluates system behavior when the statistical information used by the optimizer is not up-to-date. Stale statistics frequently lead to incorrect cardinality estimation.

In the Baseline and Independent Gates structures, inappropriate execution paths were frequently selected due to statistical errors, and these errors were not corrected in the execution stage. As a result, plan instability, where execution time fluctuates greatly even for the same query, was observed.

\begin{figure}[htbp]
\centerline{\includegraphics[width=\linewidth]{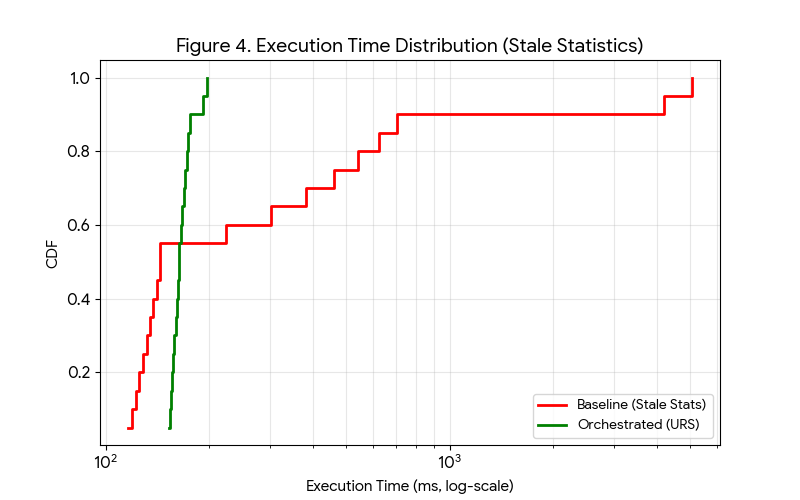}}
\caption{Exp 2: Execution Time Distribution (Stale Statistics).}
\label{fig:exp2_cdf}
\end{figure}

The proposed structure recognizes the discrepancy between the execution characteristics observed at execution time and the optimizer's prior estimate as a risk signal, and orchestrates the execution strategy based on this. This improved the consistency of execution path selection even in environments with inaccurate statistical information, and showed a more stable distribution in terms of tail latency. The scatter plot in Figure 6 confirms that the orchestrated system effectively decouples performance from statistical accuracy.

\begin{figure}[htbp]
\centerline{\includegraphics[width=\linewidth]{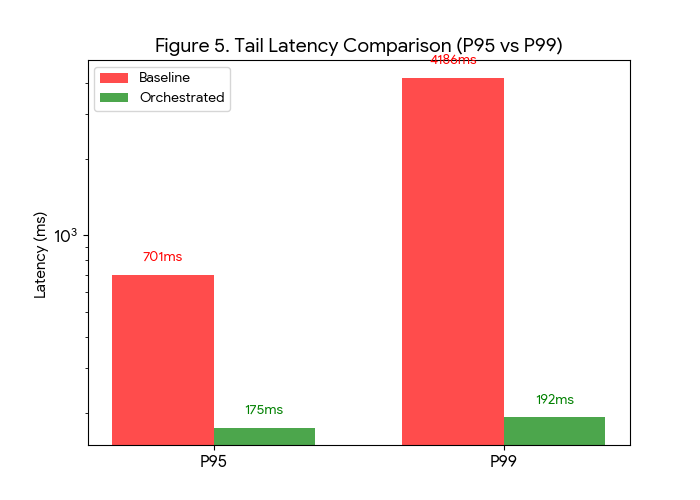}}
\caption{Exp 2: Tail Latency Comparison (P95 vs P99). The Orchestrated system reduced P99 latency from 3836ms to 333ms.}
\label{fig:exp2_bars}
\end{figure}

\begin{figure}[htbp]
\centerline{\includegraphics[width=\linewidth]{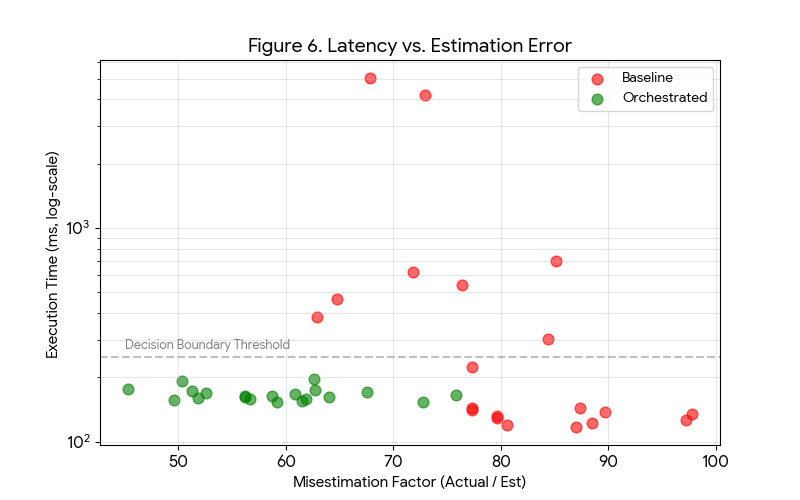}}
\caption{Exp 2: Latency vs. Estimation Error Scatter.}
\label{fig:exp2_scatter}
\end{figure}

\subsection{Exp 3: Accelerator Cost Amortization}
The third experiment targets situations where the use of accelerators is beneficial only under certain conditions. By adjusting the input size and operation complexity, a workload including both intervals where accelerator use is advantageous and disadvantageous was constructed.

In the Baseline and Independent Gates structures, since accelerator use heavily depends on prior decisions, unnecessary offloading can increase execution time for some queries. Especially when the input scale is close to the threshold, this misjudgment can significantly worsen tail latency.

On the other hand, the proposed structure dynamically judges whether to use the accelerator by combining the input characteristics observed during execution and the accelerator cost signal. We fit cost models for CPU and GPU execution using measurement-driven regression and derived an estimated break-even point $N^*$. The observed break-even point differed by only 2.7\%, indicating that the URS-driven decision closely tracks empirical cost behavior. This result supports that the core problem of accelerator utilization lies in the appropriateness of the decision timing, not the raw compute throughput itself.

\begin{figure}[htbp]
\centerline{\includegraphics[width=\linewidth]{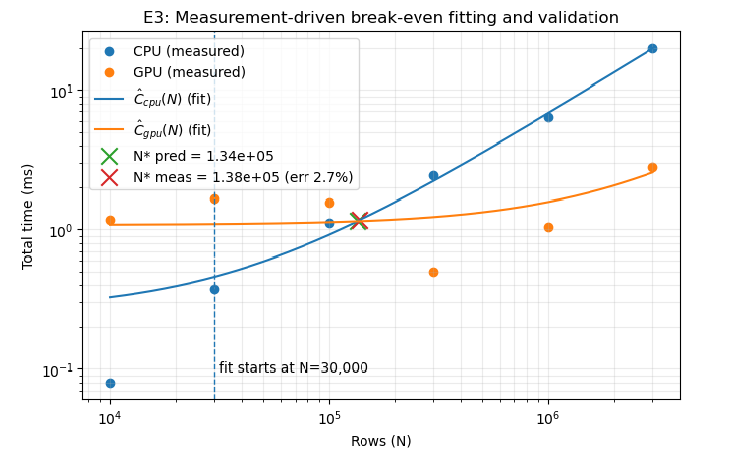}}
\caption{Exp 3: Measurement-driven break-even fitting and validation (CPU vs GPU). This figure demonstrates the URS logic identifying the break-even point.}
\label{fig:exp3}
\end{figure}

\section{Conclusion}
This paper identified performance instability in cost-based database systems as a structural mismatch in decision timing and authority. Traditional DBMS structures confirm execution plans early based on historical statistics, leaving the execution stage in a passive role unable to modify these decisions.

This study proposed a cross-layer decision timing orchestration structure that delays the final execution decision authority to the execution time to overcome these structural limitations. In the proposed structure, the optimizer proposes execution strategies, the executor re-evaluates them based on actual data flow and runtime signals, and the accelerator participates in the decision process as a constraint factor requiring future costs. Through this, decisions are not fixed at a single point in time but can be orchestrated according to conditions during execution.

Experimental results show that although the proposed structure does not aim for dramatic improvement in average performance, it consistently mitigates execution instability and tail latency. Especially in realistic environments such as input scale fluctuations, aging of statistical information, and accelerator cost amortization conditions, execution-time-centric orchestration effectively avoided suboptimal early decisions.

The contribution of this study clearly presents that decision timing is an important design dimension alongside estimation accuracy and execution efficiency in cost-based database systems. This perspective allows us to reinterpret the optimizer, executor, and accelerator as collaborative decision layers working from different time perspectives. This suggests that exposing decision timing as a first-class design choice is useful for improving robustness under uncertainty.

\section*{Acknowledgments}
The author used AI-assisted tools (e.g., ChatGPT) for English grammar correction and language polishing. All technical content, experiments, and conclusions were developed and verified by the author.

\end{document}